\documentclass[prc,amsmath,twocolumn,showpacs,superscriptaddress]{revtex4}
\usepackage{dcolumn}
\usepackage{graphicx}

\begin{document}

\title{Coupling and higher-order effects in the $^{12}$C($d$,$p$)$^{13}$C and
$^{13}$C($p$,$d$)$^{12}$C reactions}

\author{F. Delaunay}
\email{delaunay@nscl.msu.edu}
\affiliation{National Superconducting Cyclotron Laboratory, 
Michigan State University, East Lansing, Michigan 48824}
\author{F. M. Nunes}
\author{W. G. Lynch}
\author{M. B. Tsang}
\affiliation{National Superconducting Cyclotron Laboratory, 
Michigan State University, East Lansing, Michigan 48824}
\affiliation{Department of Physics and Astronomy, Michigan State University,
East Lansing, Michigan 48824}

\date{\today}

\begin{abstract}
Coupled channels calculations are performed for the $^{12}$C($d$,$p$)$^{13}$C
and $^{13}$C($p$,$d$)$^{12}$C reactions between 7 and 60 MeV to study the
effect of inelastic couplings in transfer reactions.
The effect of treating transfer beyond Born approximation is also addressed.
The coupling to the $^{12}$C 2$^+$ state is found to change the peak
cross-section by up to 15 \%. Effects beyond Born approximation lead to a
significant renormalization of the cross-sections, between 5 and 10 \% for
deuteron energies above 10 MeV, and larger than 10 \% for lower energies. We
also performed calculations including the remnant term in the transfer
operator, which has a small impact on the $^{12}$C($d$,$p$)$^{13}$C(g.s.) and
$^{13}$C($p$,$d$)$^{12}$C(g.s.) reactions. Above 30 MeV deuteron energy, the
effect of the remnant term is larger than 10 \% for the
$^{12}$C($d$,$p$)$^{13}$C(3.09 MeV) reaction and is found to increase with
decreasing neutron separation energy for the 3.09 MeV state of $^{13}$C. This
is of importance for transfer reactions with weakly bound nuclei.
\end{abstract}

\pacs{25.45.Hi, 24.10.Eq}

\maketitle

\section{Introduction}

Single-nucleon transfer reactions, that probe the degrees of freedom of single
particles, have been extensively used to study the structure of stable nuclei.
The analysis of such reactions provides the angular momentum transfer
\cite{Satchler83}, which gives information on the spin and parity of the final
state. The sensitivity of the cross-sections to the single-nucleon components
allows for the extraction of spectroscopic factors. These observables quantify
the overlaps between nuclear states and can be used to deduce occupancies of
single-particle orbitals in nuclei \cite{Satchler83}.
Thanks to recent developments in radioactive beam production and in the
detection of light charged particles, transfer reactions can now be used to
investigate the single-particle structure of exotic nuclei.
Experimental programs on transfer reactions with radioactive beams have indeed
been initiated by several teams \cite{Rehm98,Fortier99,DeSereville03}. More
recently, nuclear knockout reactions \cite{Hansen03} were shown to be another
useful tool to extract spectroscopic factors for loosely bound nuclei. The
recent indications of reduced occupancies of single-particle states
\cite{Hansen03,Kramer01,Gade04} reveal that reliable measurements of
spectroscopic factors in exotic nuclei are highly desirable.

The analysis of transfer reactions most frequently relies on the Distorted
Waves Born Approximation \cite{Satchler83}. This method presents uncertainties
due to (i) the sensitivity to the optical model parameters, (ii) the
sensitivity to the single-particle parameters and (iii) the assumptions
behind the DWBA formalism itself. The sensitivity of DWBA cross-sections to
optical model parameters has been the object of a recent investigation in the
case of the $^{12}$C($d$,$p$)$^{13}$C and $^{13}$C($p$,$d$)$^{12}$C reactions
\cite{Liu04}.
In that work, a reanalysis of existing data between 4 and 60 MeV deuteron
energy was performed using DWBA with consistent input parameters over the
whole energy range.
In particular, potentials for the deuteron and proton channels were obtained
from global parameterisations. This was shown to reduce the variations of the
measured spectroscopic factors as a function of incident energy. The smallest
variations were obtained by using adiabatic potentials for the deuteron
channel, following the prescription by Johnson and Soper \cite{Johnson70}, in
order to take deuteron breakup into account.
With that procedure, the average spectroscopic factor for
the ground state to ground state transition in $^{12}$C($d$,$p$)$^{13}$C and
$^{13}$C($p$,$d$)$^{12}$C reactions between 12 and 60 MeV was found to be
0.61 $\pm$ 0.09, the error being the r.m.s. variation over the energy
\cite{Liu04}. This result is in excellent agreement with the prediction of
the pioneering shell-model calculations of Cohen and Kurath for $p$-shell
nuclei \cite{Cohen67}. This success could be seen as an indication that
calculations similar in form to DWBA but using adiabatic deuteron potentials
give a good description of the ($d$,$p$) and ($p$,$d$) reaction mechanisms.
However, in \cite{Liu04} no direct checks of effects beyond these
approximations were done.

DWBA assumes that elastic scattering is the dominant process in the entrance
and exit channels \cite{Satchler83}. The relative motion of the reaction
partners in each channel is thus approximated by a distorted wave describing
elastic scattering. The transfer operator is assumed to be weak, so that the
transfer transition amplitude can be treated using Born approximation
\cite{Satchler83}.
In the case of the ($d$,$p$) and ($p$,$d$) reactions, the description of the
data is generally improved by taking into account the effect of deuteron
breakup using an adiabatic deuteron potential \cite{Harvey71,Wales76}, as
already mentioned. Such an adiabatic potential is used in place of the
optical potential to generate the distorted wave for the deuteron channel,
whereas the rest of the calculation is the same as in DWBA.
This procedure will be referred to as the ADW (Adiabatic Distorted Wave)
method in the following. In that framework, the process is still assumed to
be elastic with respect to the target.

This paper reports on coupled channels calculations performed for the
$^{12}$C($d$,$p$)$^{13}$C and $^{13}$C($p$,$d$)$^{12}$C reactions to test the
effect of target excitation in entrance and exit channels.
The effect of treating transfer beyond Born approximation was also addressed.
Usually, in the analysis of ($d$,$p$) and ($p$,$d$) reactions, the transfer
operator is replaced by the neutron-proton interaction alone. We checked the
effect of this approximation by performing calculations with a more complete
transfer operator.

In most coupled channels calculations, the optical potentials are adjusted so
that the predicted elastic scattering computed with the couplings reproduces
elastic scattering data. This procedure is not applicable to the present work,
because we used adiabatic deuteron potentials which are not meant to describe
elastic scattering.
We are interested in a systematic estimation of the effect of couplings and
not in performing the most precise calculation for the transfer cross-section
at a given energy. The latter would require detailed comparisons between
measured and calculated cross-sections for all the included channels and fine
adjustments of potentials and coupling strengths. Estimation of the magnitude
of the errors due to the neglect of couplings in ADW, however, does not depend
on such details. The main goal of our work is to produce such an estimation,
in order to provide guidelines for the analysis of transfer reactions with
stable or weakly bound nuclei.

\section{Inelastic couplings}

\begin{figure}[ht]
\includegraphics[width=9cm]{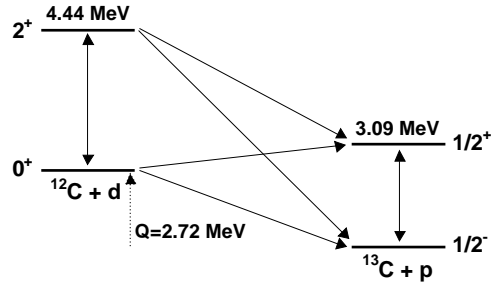}
\caption{Coupling scheme for the $^{12}$C($d$,$p$)$^{13}$C reaction.}
\label{CouplScheme}
\end{figure}

We performed coupled channels calculations for the $^{12}$C($d$,$p$)$^{13}$C
and $^{13}$C($p$,$d$)$^{12}$C reactions, including transfer routes through the
4.44 MeV 2$^+$ state of $^{12}$C and the 3.09 MeV 1/2$^+$ state of $^{13}$C.
The calculations were performed with the code FRESCO \cite{Thompson88}, in the
framework of the Coupled Channels Born Approximation (CCBA) \cite{Satchler83}.
We used adiabatic potentials for the deuteron channel, thus our treatment
differs from usual CCBA calculations. We have shown that
the adiabatic approximation for deuteron breakup can be introduced in the
coupled channels treatment of inelastic target excitations \cite{Delaunay}.
In the case of transfer, this leads to coupled equations and a transfer
transition amplitude similar in form to the usual CCBA \cite{Delaunay}. For
the sake of clarity, our treatment will be dubbed as ACC (for Adiabatic
Coupled Channels) in the following. In this framework, target excitations in
the entrance and exit channels are included explicitly and treated to all
orders by solving coupled equations, whereas the transfer is still treated
using Born approximation.
The coupling scheme of Fig. \ref{CouplScheme} summarises the transfer routes
and excitations included in the calculations, in the case of the
$^{12}$C($d$,$p$)$^{13}$C reaction.

The 2$^+$ state of $^{12}$C was assumed to be a collective excitation of
rotational nature. The coupling between this state and the $^{12}$C ground
state was included by deforming the $d$+$^{12}$C potential. The deformation
length was obtained from the adopted experimental value of
$39.7\pm3.3$~$e^2$fm$^4$ for the B(E2) of $^{12}$C \cite{Raman01}.
Using the collective model with a value of 1.25 $\times$ 12$^{1/3}$ fm for
the radius of $^{12}$C, the deformation length was found to be
$\beta_2 R = -1.54$ fm. The negative sign was chosen following experimental
indications of an oblate deformation \cite{Vermeer83}.

The 1/2$^+$ first excited state of $^{13}$C was assumed to be a
single-particle E1 transition. The single-neutron excitation in $^{13}$C due
to the interaction with the proton was performed by using a folded coupling
potential in place of the $^{13}$C+$p$ optical potential. This coupling
potential was taken as the sum of a $^{12}$C+$p$ potential and a
neutron-proton interaction. The $^{12}$C+$p$ potential was built from the
JLM interaction. The neutron-proton part was a gaussian interaction, with
parameters taken from \cite{Assuncao04} and reproducing the r.m.s. radius and
binding energy of the deuteron.

Spectroscopic amplitudes for the overlaps of the $^{13}$C states with the
$^{12}$C states were obtained from shell-model calculations, performed
in the $psd$ model space using the code OXBASH \cite{Oxbash}, with the
interaction named PSDMK in the OXBASH nomenclature. This interaction is
composed of three parts: (i) the Cohen and Kurath (8-16)POT potential
representation for the $p$-shell \cite{Cohen65}, (ii) the $sd$-shell
interaction of Preedom and Wildenthal \cite{Preedom72}, and (iii) the
$p$-$sd$ cross-shell interaction of Millener and Kurath \cite{Millener75}.
The 0$^+$ ground state and 2$^+$ state of $^{12}$C and the 1/2$^-$
ground state of $^{13}$C were assumed to be 0 $\hbar \omega$ states, whereas
the 1/2$^+$ state of $^{13}$C was assumed to be a 1 $\hbar \omega$ state. The
resulting spectroscopic amplitudes are summarized in Table \ref{TabAmp}. As a
check, we also computed the B(E2) reduced transition probability using the
$^{12}$C $0^+$ and $2^+$ shell-model wave functions. We obtained a value of
50.3 $e^2$fm$^4$, close to the adopted experimental value. The sign of the
computed E2 transition matrix element was also found to be in agreement with
the sign of the deformation length of $^{12}$C.

\begin{table}[htb]
\begin{center}
\begin{tabular}{lcccc}
\hline
\hline
$^{12}$C($J^\pi$)$\otimes$($n\ell j$)$_\nu$ & & $^{13}$C($\frac{1}{2}^-$,g.s.)
& $^{13}$C($\frac{1}{2}^+$,3.09 MeV) \\
\hline
$0^+_{gs}\otimes 1p_{1/2}$    & & $-0.7755$      &           \\
$2^+_1\otimes 1p_{3/2}$       & & $1.0592$       &           \\
$0^+_{gs}\otimes 2s_{1/2}$    & &                & $-0.9282$ \\
$2^+_1\otimes 1d_{5/2}$       & &                & $-0.3089$ \\
\hline
\hline
\end{tabular}
\end{center}
\caption{Spectroscopic amplitudes for the $\langle ^{12}$C$ \otimes (n\ell
j)_\nu| ^{13}$C$ \rangle$ overlaps, as computed from the shell-model
calculations described in the text.}
\label{TabAmp}
\end{table}

The single-neutron radial form factors were computed in a Woods-Saxon well
with a standard geometry ($r=1.25$ fm and $a=0.65$ fm, as used in
\cite{Liu04}). Each form factor was calculated separately by adjusting the
depth of the potential in order to obtain the experimental binding energy,
and then normalised by the relevant shell-model spectroscopic amplitude.

Apart from the aspects of the calculations specific to the coupled channels
approach, our work has a few minor differences with the work of Liu et al.
\cite{Liu04}:
\begin{itemize}
\item our calculations were performed in exact finite range, using
the Reid Soft Core interaction for the neutron-proton interaction to compute
the deuteron ground state wave function and in the transfer operator. The
representation of the transition amplitude \cite{Satchler83} was chosen so as
to obtain the neutron-proton interaction as the main transfer operator. We
thus adopted the post representation for $^{12}$C($d$,$p$)$^{13}$C and the
prior representation for $^{13}$C($p$,$d$)$^{12}$C~;
\item we included a spin-orbit interaction in the single-particle potential
used to compute the neutron form factors for $^{13}$C, with a depth of
$V_{SO}=6$ MeV~;
\item we did not include any non-locality corrections in the calculations,
since these corrections are very complicated within the coupled channels
formalism.
\end{itemize}
These modifications are expected to have the same effect on ADW and ACC
calculations, and therefore should not change our conclusions on couplings.
All other aspects of the calculations were chosen to be the same as in the
study by Liu et al. \cite{Liu04}.
We therefore used the same adiabatic potentials for the deuteron channels. The
$n$+$^{12}$C and $p$+$^{12}$C optical potentials used to build these potentials
as well as the $p$+$^{13}$C optical potentials were obtain from the JLM
microscopic nucleon-nucleon interaction. Further details on these potentials
can be found in the paper by Liu et al. \cite{Liu04}.

\begin{figure}[htb]
\begin{center}
\includegraphics[width=4.2cm]{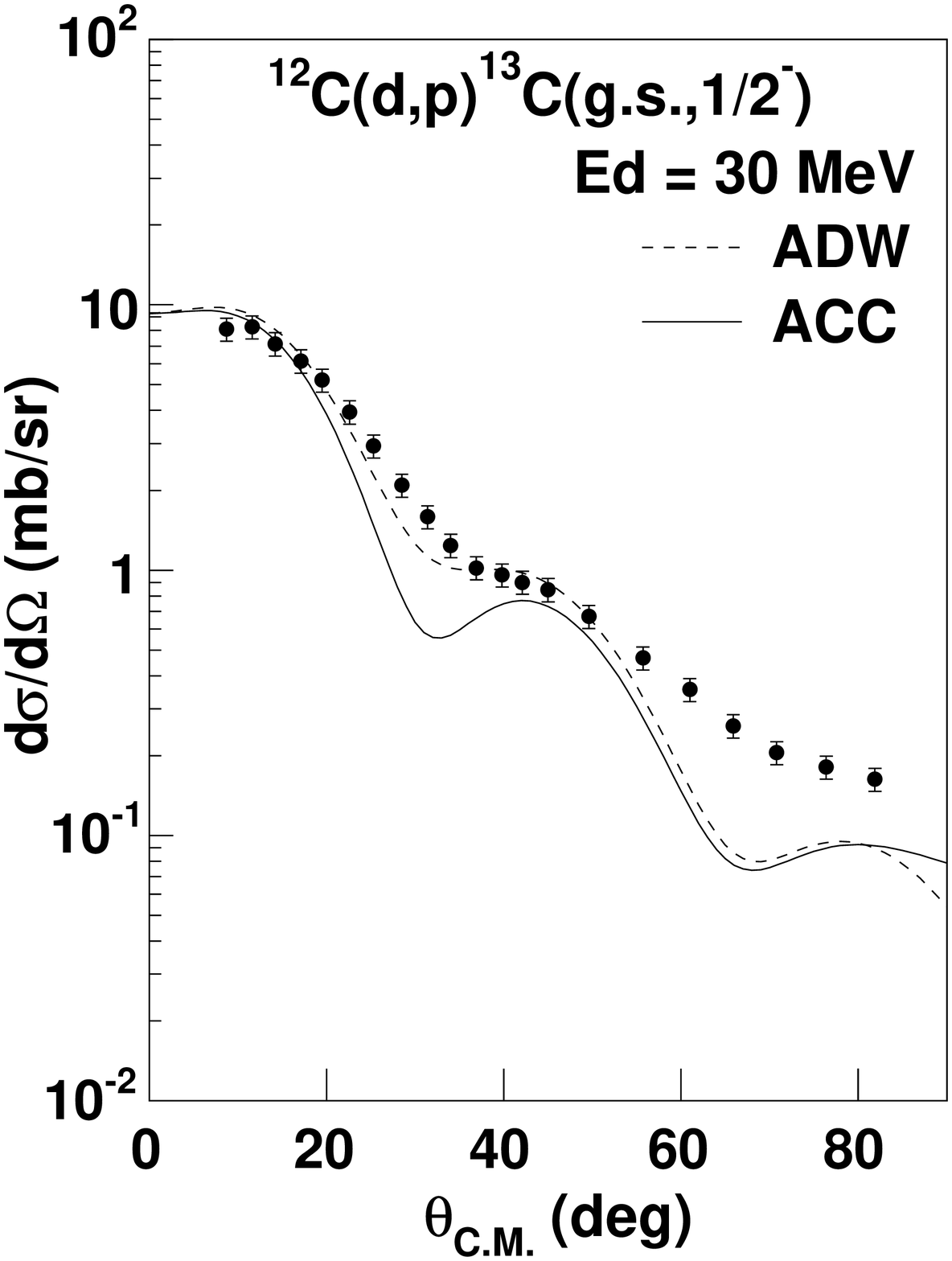}
\includegraphics[width=4.2cm]{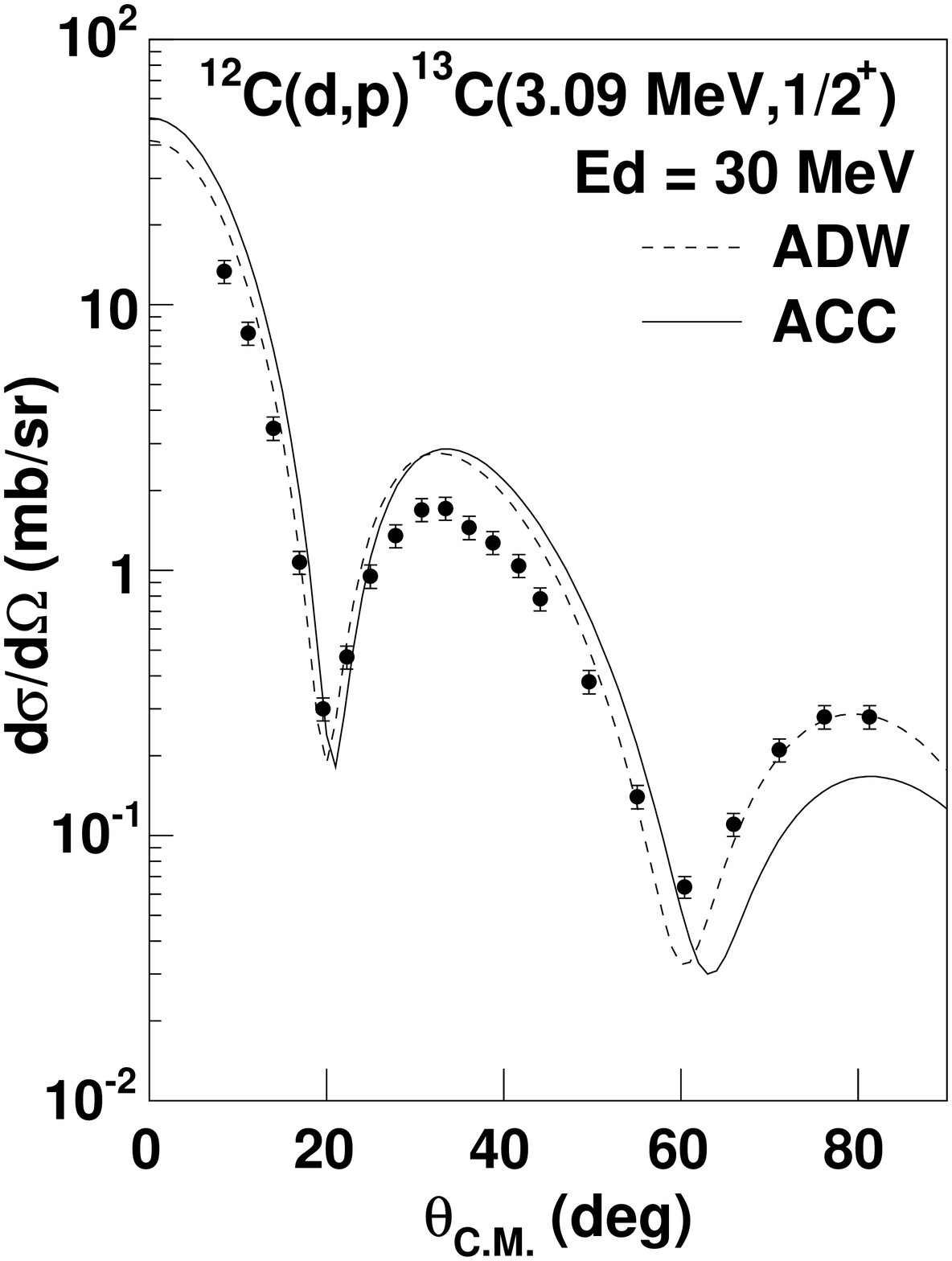}
\end{center}
\caption{Experimental, ADW and ACC angular distributions for the
$^{12}$C($d$,$p$)$^{13}$C reaction leading to the ground state and first
excited state of $^{13}$C, at 30 MeV incident energy.}
\label{AngDistCCBA}
\end{figure}

In Fig. \ref{AngDistCCBA}, we present results of ADW and ACC calculations
for the $^{12}$C($d$,$p$)$^{13}$C reaction leading to the ground and first
excited states of $^{13}$C at 30 MeV deuteron energy, as compared to the data
of Ohnuma et al. \cite{Ohnuma86}. ADW calculations were performed using the
same parameters as for ACC calculations, but without inelastic couplings in
the entrance and exit channels. Therefore, transfer paths through an excited
state of $^{12}$C and/or $^{13}$C are absent from the ADW calculations.
These calculations were not normalised to the data and no attempt was made to
improve the description of the data. Nevertheless, Fig. \ref{AngDistCCBA}
shows that both ADW and ACC calculations can reproduce the magnitude of the
forward angle cross-sections for the $^{12}$C($d$,$p$)$^{13}$C reaction.
When comparing ACC and ADW angular distributions, one sees that the couplings
change both the shape and amplitude in the peak region, where spectroscopic
factors are usually extracted by normalizing calculated cross-sections to the
data. At 30 MeV, the change in the cross-section at the angle corresponding
to the ADW peak is - 3~\% for the transition to the ground state of $^{13}$C
and + 22~\% for the transition to the first excited state of $^{13}$C.

From Fig. \ref{AngDistCCBA}, it also appears that the description of the data
is not improved by doing ACC calculations.
Better agreement might be obtained if the deuteron potential was constrained
to fit the deuteron elastic scattering data.
Here, the deuteron potential in the ACC calculations was kept the same as in
the ADW calculations. The couplings to excited states modified the
$d$+$^{12}$C(g.s.) and $p$+$^{13}$C(g.s.) relative wave functions, which
could have an impact on the transfer cross-sections. In order to clarify this
effect, we performed another series of ACC calculations with one-way couplings
only, i.e. without back couplings from the excited states of the $^{12}$C and
$^{13}$C to their respective ground-states. This means that no couplings were
taken into account in the coupled equations describing the
relative motions of the $^{12}$C(g.s.)+$d$ and $^{13}$C(g.s.)+$p$ systems.
Thus, in these latter calculations, the elastic wave functions were not
modified in the entrance channel nor in the exit channel when going from ADW
to ACC. We found that the effects of one-way couplings are of the same
magnitude and sign as with the full couplings.

Describing deuteron elastic scattering and deuteron breakup at the same time
would require to treat the deuteron breakup in an explicit way using more
complicated models, such as the Continuum Discretised Coupled Channels (CDCC)
method. Such calculations were performed recently by Keeley et al.
\cite{Keeley04} for the $^{12}$C($d$,$p$)$^{13}$C reaction at 15 and 30 MeV
deuteron energy, also including the excitation of the $^{12}$C 2$^+$ state.
Although these authors used a more realistic description of the deuteron
breakup through CDCC and used deuteron elastic scattering data to adjust the
$d$+$^{12}$C potentials, the quality of the agreement between the data and
their calculations is at the same level as for our ACC calculations.
It is not the purpose of this work to provide a good fit to the data but
rather to probe whether the neglect of couplings to excited states in the ADW
approach leads to a significant error in the extracted spectroscopic factors.

\begin{figure}[htb]
\begin{center}
\includegraphics[width=8.5cm]{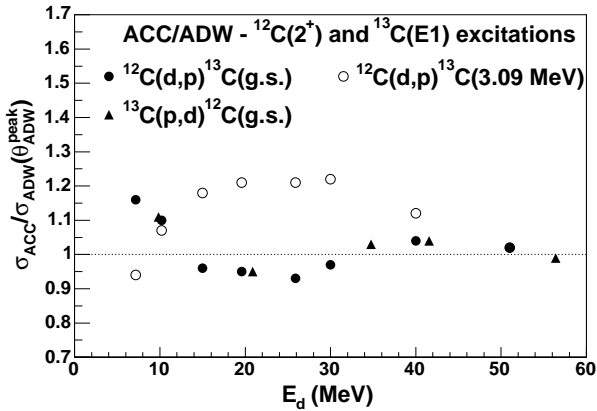}
\end{center}
\caption{Ratio of the ACC to ADW cross-sections taken at the angle of the
ADW peak, for the $^{12}$C($d$,$p$) reaction leading to the ground state and
first excited state of $^{13}$C and for the $^{13}$C($p$,$d$) reaction
leading to the ground state of $^{12}$C. ACC calculations include excitation
of both the $^{12}$C 2$^+$ and $^{13}$C $1/2^+$ states.}
\label{VareCCBA}
\end{figure}

In Fig. \ref{VareCCBA}, we show the ratio of ACC to ADW cross-sections at the
ADW peak as a function of deuteron energy, for the $^{12}$C($d$,$p$)$^{13}$C
reaction leading to the ground and first excited states of $^{13}$C, and for
the $^{13}$C($p$,$d$)$^{12}$C reaction leading to the $^{12}$C ground state.
For the ground-state to ground-state transitions, the effect seems to be
increasing with decreasing energy and shows an oscillatory behavior. We also
note that couplings have identical effects on the
$^{12}$C($d$,$p$)$^{13}$C(g.s.) and $^{13}$C($p$,$d$)$^{12}$C(g.s.) reactions,
as expected from the detailed balance principle.
For deuteron energies above 10 MeV, the effect on the ground state to ground
state transitions is of the order of 5 \%. The effect is much stronger in
the case of the $^{12}$C($d$,$p$)$^{13}$C($1/2^+$,3.09 MeV) reaction and leads
to a change in the peak cross-section larger than 15 \% for deuteron energies
between 15 and 30 MeV.
This shows that the effect of couplings can be large and depends on the final
state and on the incident energy. In other words, a small coupling effect at
a particular energy does not mean that the effect is small at all energies.

Previous CCBA analysis for the $^{12}$C($d$,$p$)$^{13}$C reaction to positive
parity states in $^{13}$C were performed at 15 MeV \cite{Tanifuji82} and 30
MeV \cite{Ohnuma86} deuteron energy, with optical potentials for the deuteron
channel. At 15 MeV, the effect on the $1/2^+$ state of $^{13}$C due to the
coupling to the $^{12}$C $2^+$ state is smaller that the one observed in
the present work. This might be attributed to different choices for the
deuteron+$^{12}$C potential. At 30 MeV, the effect from the $2^+$ is
comparable to our observations. Those two previous studies did not address
the effect of couplings on the transfer to the $^{13}$C ground state.

\begin{figure}[htb]
\begin{center}
\includegraphics[width=8.5cm]{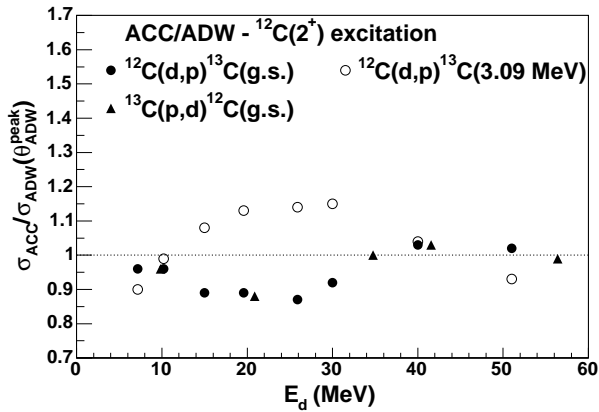}
\end{center}
\caption{Ratio of the ACC to ADW cross-sections taken at the angle of the
ADW peak, for the $^{12}$C($d$,$p$) reaction leading to the ground state and
first excited state of $^{13}$C and for the $^{13}$C($p$,$d$) reaction leading
to the ground state of $^{12}$C. ACC calculations include the excitation of
the $^{12}$C 2$^+$ state only and neglect the E1 coupling to the $^{13}$C
3.09 MeV state.}
\label{VareCCBA2}
\end{figure}

The assumption of a single-particle E1 excitation for the $^{13}$C 3.09 MeV
state can be tested by comparing the experimental B(E1) for the corresponding
transition to the B(E1) calculated using the neutron form factors and the
shell-model spectroscopic amplitudes. The computed B(E1) is 0.28 Weisskopf
units, whereas the experimental value is $0.047 \pm 0.010$ W. u.
\cite{Ajzenberg91}. This strong overestimation is due to the fact that a
significant fraction of the E1 strength is actually located at higher
excitation energy, in the region of the giant dipole resonance, which was not
taken into account here. Consequently, the effect of the excitation of the
$^{13}$C first excited state on the transfer is overestimated.
To clarify the uncertainty concerning this E1 excitation, we performed ACC
calculations with the excitation of the $^{12}$C 2$^+$ state alone. The
ratios of these ACC cross-sections to the ADW cross-sections are
shown on Fig. \ref{VareCCBA2}. We see that the overall effect of the
excitation of the $^{12}$C 2$^+$ state is similar to the effect obtained by
including both the 2$^+$ and E1 excitations. Again, ACC to ADW ratios for the
$^{12}$C($d$,$p$)$^{13}$C(g.s.) and $^{13}$C($p$,$d$)$^{12}$C(g.s.) reactions
show identical behaviors, which implies that the coupling to the 2$^+$ state
has the same effect whether it occurs in the entrance or exit channel. Between
15 and 30 MeV, switching off the E1 excitation changes the peak cross-section
by about 5 \%. Below 15 MeV, this change is of the order of 10 \%. For the
$^{12}$C($d$,$p$)$^{13}$C(3.09 MeV) reaction, the effect of removing the E1
excitation in $^{13}$C amounts to a 5 \% change in the cross-section, at all
energies. The excitation of the $^{12}$C 2$^+$ state alone already has an
important effect on transfer cross-sections, which are changed by about 15 \%
at 25 MeV, for the three reactions studied. Since we used the experimental
B(E2) value to determine the intensity of the coupling between the ground
state and $2^+$ state of $^{12}$C, this excitation is not subject to the same
uncertainty as the coupling to the $^{13}$C 3.09 MeV state.

\section{The remnant term}

In the post representation of DWBA for the $^{12}$C($d$,$p$)$^{13}$C reaction,
the transfer operator is
\cite{Satchler83}
\begin{equation}
W_{^{13}Cp} = V_{^{13}Cp} - U_{^{13}Cp},
\end{equation}
where $V_{^{13}Cp}$ is the full effective interaction between the two nuclei
in the exit channel, and $U_{^{13}Cp}$ is the optical potential used to
generate the distorted wave in the exit channel. If one assumes that the full
interaction is the sum of two-body interactions, $W_{^{13}Cp}$ can be
decomposed into
\begin{equation}
W_{^{13}Cp} = V_{np} + V_{^{12}Cp} - U_{^{13}Cp},
\end{equation}
where $V_{^{12}Cp}$ is the \textit{core-core interaction} and
$V_{^{12}Cp} - U_{^{13}Cp}$ is called \textit{remnant term} or
\textit{indirect interaction}. The remnant term is usually neglected in DWBA
and ADW calculations, leaving the neutron-proton interaction as the only
transfer operator. This assumption is questionable, especially in the case of
light and/or exotic nuclei for which the interaction $V_{Ap}$ could differ
significantly from the $V_{(A-1)p}$ interaction.

\begin{figure}[htb]
\begin{center}
\includegraphics[width=8.5cm]{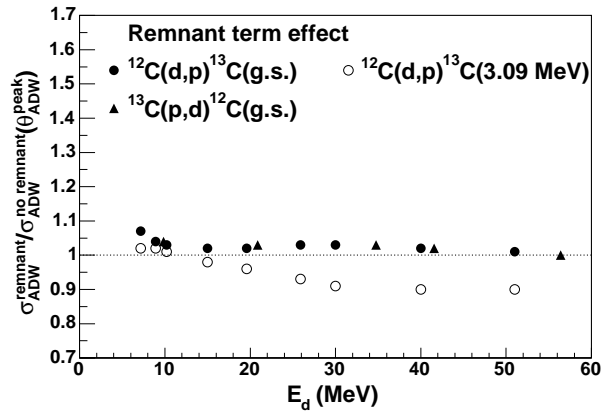}
\end{center}
\caption{Ratio of the ADW cross-sections calculated with and without remnant
term, as a function of deuteron energy.}
\label{VareRemnant}
\end{figure}

We performed a series of ADW calculations for the $^{12}$C($d$,$p$)$^{13}$C
and $^{13}$C($p$,$d$)$^{12}$C reactions, including the remnant term in the
transfer operator. The core-core interactions were assumed to be complex, as
indicated by Satchler \cite{Satchler83}, and were approximated by $^{12}$C+$p$
optical potentials obtained using the JLM interaction. In the transfer
calculations, we included both real and imaginary parts of the remnant term.
Ratios of ADW peak cross-sections with and without remnant term are displayed
on Fig. \ref{VareRemnant}, as a function of deuteron energy. Here again,
effects on the $^{12}$C($d$,$p$)$^{13}$C(g.s.) and
$^{13}$C($p$,$d$)$^{12}$C(g.s.) are found to be identical. The effect of the
remnant term on these ground state to ground state transitions is smaller than
5 \%, except for the lowest energies, i.e. at about 8 MeV. In the case of the
($d$,$p$) reaction to the $^{13}$C first excited state, inclusion of the
remnant term lowers the peak cross-section. This effect increases with energy
and seems to saturate at 10 \% for energies over 30 MeV. Effects of the
remnant term on ACC cross-sections are very similar to the effects on ADW
calculations.

\begin{figure}[htb]
\begin{center}
\includegraphics[width=7.3cm]{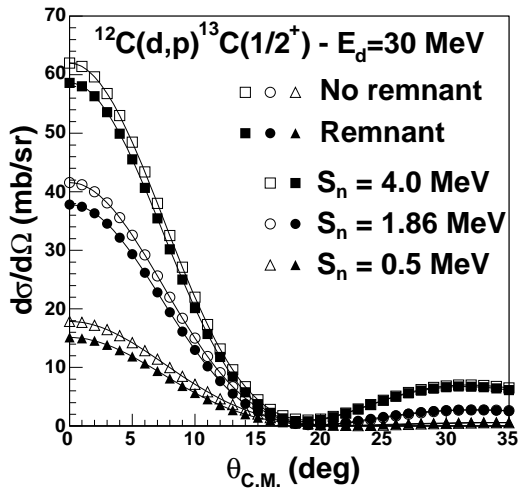}
\end{center}
\caption{ADW angular distributions for the $^{12}$C($d$,$p$)$^{13}$C($1/2^+$)
reaction at 30 MeV, with and without remnant term, and with different neutron
separation energies for the final state. The lines are displayed to guide the
eye.}
\label{RemnantBE13C1}
\end{figure}

The greater sensitivity of the $^{12}$C($d$,$p$)$^{13}$C($1/2^+$,3.09 MeV)
reaction to the remnant term is particularly interesting. The neutron
separation energy of the 3.09 MeV state in $^{13}$C is only 1.86 MeV, whereas
it is 4.95 MeV for the ground state. In order to test the relation between
the separation energy and the effect of the remnant term, we performed ADW
calculations for the $^{12}$C($d$,$p$)$^{13}$C reaction at 30 MeV, with and
without remnant term, and with neutron separation energies artificially
modified for both states in $^{13}$C.
We calculated the transfer cross-sections to a $1/2^+$ state in $^{13}$C
($2s_{1/2}$ neutron configuration) with $S_n$ = 0.5 and 4.0 MeV and to a
$1/2^-$ state ($1p_{1/2}$ neutron configuration) with $S_n$ = 0.5 and 2.0 MeV.
The results for the transition to the $1/2^+$ state are shown on Fig.
\ref{RemnantBE13C1}. The cross-section decreases when the neutron separation
energy decreases, whereas the absolute effect of the remnant term on the peak
cross-section is roughly constant. Therefore, the relative effect of the
remnant term increases when the neutron separation energy decreases. For a
binding energy of 0.5 MeV, inclusion of the remnant term reduces the ADW peak
cross-section by 15 \%. For  the transition to the $1/2^-$ state, the effect
of the remnant term is smaller than 5 \% for $S_n = 4.95$ MeV, and decreases
with decreasing binding energy. At $S_n$ = 0.5 MeV, the effect on the peak
cross-section is smaller than 1 \%. Clearly, the effect of the remnant term
depends on the properties of the populated neutron state. In particular, we
find that large corrections are to be expected when transferring to loosely
bound neutron $s_{1/2}$ orbitals, which play an important role in the
appearance of nuclear halo states \cite{Hansen95}.
For the interaction of the proton and the $^{13}$C, we used optical potentials
built from the JLM interaction using the ground state $^{13}$C matter density
\cite{Liu04}.
When using more realistic optical potentials for loosely bound systems
(typically more diffuse), the remnant term effect will be even larger than
our estimate.

\section{CRC calculations}

\begin{figure}[htb]
\begin{center}
\includegraphics[width=8.5cm]{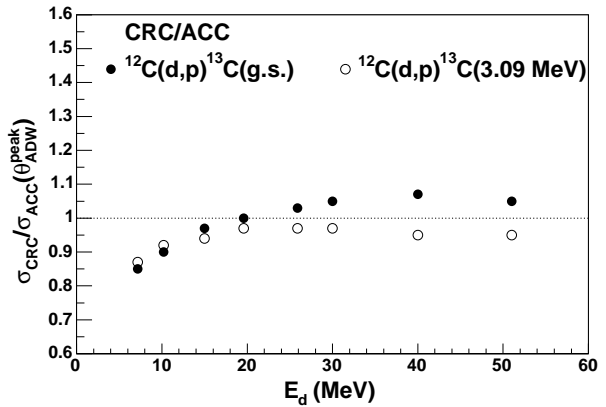}
\end{center}
\caption{Ratio of the CRC to ACC cross-sections at the ADW peak angle, for
the $^{12}$C($d$,$p$)$^{13}$C(g.s.) and $^{12}$C($d$,$p$)$^{13}$C(3.09 MeV)
reactions.}
\label{VareCRC}
\end{figure}

If the transfer interaction is strong enough, it can induce multiple transfers
between the initial and final states, in the forward direction (i.e. from the
entrance channel to the exit channel) and in the backward direction (i.e.
from the exit channel to the entrance channel). One then needs to go beyond
Born approximation to compute the transfer transition amplitude.
To check the validity of Born approximation for the $^{12}$C($d$,$p$)$^{13}$C
and $^{13}$C($p$,$d$)$^{12}$C reactions, we performed Coupled Reaction
Channels calculations.
In this method, the transfer is treated to an arbitrarily high order. This
order is the number of times the nucleon is transferred between the entrance
and exit partitions.
When more than one transfer step is performed, one needs to take into account
the non-orthogonality term \cite{Satchler83} between the entrance and exit
partitions.
At each step in CRC, all couplings are treated by solving coupled equations,
as in the CCBA method, but including also a source term due to the transfer.
The channel wave functions used in the source terms at a given step are taken
as the results of the previous step.
In the CRC calculations, we included the remnant term in the transfer
operator, as well as the non-orthogonality correction. Results are shown in
Fig. \ref{VareCRC} in terms of the ratios of CRC and ACC cross-sections as a
function of deuteron energy.
Convergence of the CRC results for the $^{12}$C($d$,$p$)$^{13}$C reaction was
reached after 8 transfer steps. In the case of the $^{13}$C($p$,$d$)$^{12}$C
reaction, convergence was difficult to obtain, due to the remnant term and/or
the non-orthogonality correction. Therefore we do not show the corresponding
results.
Relative effects of the CRC on the $^{12}$C($d$,$p$)$^{13}$C(g.s) and
$^{12}$C($d$,$p$)$^{13}$C(3.09 MeV) reactions are similar to each other.
Below 10 MeV deuteron energy, the CRC reduces the peak cross-section by 10 \%
or more. There is a global decrease with incident energy, reflecting the
absolute transfer cross-section decrease with energy for both reactions. The
transfer channel gets weaker with increasing energy, and effects beyond Born
approximation tend to decrease. At 50 MeV deuteron energy, the effect
saturates at about  5 \% for both reactions.

\section{Summary and conclusions}

We performed coupled channels calculations for the $^{12}$C($d$,$p$)$^{13}$C
and $^{13}$C($p$,$d$)$^{12}$C reactions
in order to test the effect of inelastic couplings. We studied the effect of
the excitation of the 4.44 MeV $2^+$ state in $^{12}$C and of the $^{13}$C
$1/2^+$ state at 3.09 MeV. The coupling to the $2^+$ state changes the peak
cross-section by up to 15 \%, which shows that attention should be paid to
such excitations in transfer reactions. The rather strong energy dependence
suggests that, when analysing transfer reactions, coupled channels
calculations performed to check inelastic couplings should be done at the
energies of interest. Indeed, a small effect at a particular energy does not
guarantee that the effect is small at all energies. Assuming that the $1/2^+$
state in $^{13}$C is a single-particle E1 excitation, we found that the
coupling to this state has a somewhat smaller effect on the transfer
cross-sections. Only at the lowest energies is this effect larger than 10 \%.
Considering the fact that the B(E1) reduced transition probability to this
state is overestimated, the coupling effect would probably be significantly
smaller when this excitation is taken into account in a more realistic way.

The effect of including the remnant term in the transfer operator for the
$^{12}$C($d$,$p$)$^{13}$C(g.s.) and $^{13}$C($p$,$d$)$^{12}$C(g.s.) reactions
is smaller than 5 \% for energies above 8 MeV. In the case of the
$^{12}$C($d$,$p$)$^{13}$C($1/2^+$,3.09 MeV) reaction, the effect of the
remnant term is found to be more important than for the
$^{12}$C($d$,$p$)$^{13}$C(g.s.) reaction ($\approx$ 10 \%). We find that the
relative effect of the remnant term on the transfer to the $^{13}$C $1/2^+$
state increases with decreasing neutron separation energy, whereas the
opposite happens for the $^{12}$C($d$,$p$)$^{13}$C($1/2^-$) transition.
A large contribution from the remnant term should therefore be expected for
neutron transfer reactions to or  from loosely bound $s_{1/2}$ orbitals, e.g.
neutron halo states.

Effects beyond Born approximation were also addressed by performing CRC
calculations, including the remnant term and the non-orthogonality correction.
An effect larger than 10 \% is observed for deuteron energies below 10 MeV.
Above that energy, changes in the peak cross-sections are of the order of 5 \%.
In the case of transfer reactions involving exotic nuclei, couplings to breakup
states are expected to enhance the effect of going beyond Born approximation.
However, non-orthogonality issues need to be clarified before obtaining a
quantitative result. Work along these lines is planned for the
$^{10}$Be($d$,$p$)$^{11}$Be reaction.

\begin{acknowledgments}
We wish to thank Prof. B. A. Brown, Prof. R. C. Johnson, Prof. I. J. Thompson
and Prof. J. A. Tostevin for helpful discussions. This work was supported by
the National Science Foundation under Grant No. PHY-01-10253.
\end{acknowledgments}

\end{document}